\documentclass[aps,prd,showpacs,groupedaddress,floatfix,nofootinbib]{revtex4}
\usepackage{amsmath}
\usepackage{graphicx}
\usepackage{color}

\def\beq{\begin{eqnarray}}
\def\eeq{\end{eqnarray}}

\newcommand{\LSF}{\mathrm{LSF}}

\begin{document}

\title{Collocation method for fractional quantum mechanics}
\author{Paolo Amore}
\email{paolo.amore@gmail.com}
\affiliation{Facultad de Ciencias, CUICBAS,Universidad de Colima, \\
Bernal D\'{\i}az del Castillo 340, Colima, Colima, Mexico}
\author{Francisco M. Fern\'andez}
\email{fernande@quimica.unlp.edu.ar}
\affiliation{INIFTA (Conicet, UNLP), Division Quimica Teorica, Diagonal 113 y 64 S/N, \\
Sucursal 4, Casilla de correo 16, 1900 La Plata, Argentina}
\author{Christoph P. Hofmann}
\email{christoph@ucol.mx}
\affiliation{Facultad de Ciencias, CUICBAS, Universidad de Colima,~ \\
Bernal D\'{\i}az del Castillo 340, Colima, Colima, Mexico}
\author{Ricardo A. S\'aenz}
\email{rasaenz@ucol.mx}
\affiliation{Facultad de Ciencias, CUICBAS, Universidad de Colima,~ \\
Bernal D\'{\i}az del Castillo 340, Colima, Colima, Mexico}

\begin{abstract}
We show that it is possible to obtain numerical solutions to quantum mechanical problems involving 
a fractional Laplacian, using a collocation approach based on Little Sinc Functions (LSF), which
discretizes the Schr\"odinger equation on a uniform grid. The different boundary conditions
are naturally implemented using sets of functions with the appropriate behavior. Good convergence
properties are observed. A comparison with results based on a WKB analysis is performed.
\end{abstract}

\maketitle

\section{Introduction}

There has recently been great interest in what is called fractional
quantum mechanics. Laskin \cite{L00a,L00b,L00c} derived a fractional
Schr\"{o}dinger equation from a fractional version of the path integral. The
nature of the fractional quantum mechanics is determined by the L\'{e}vy
index $0<\alpha \leq 2$ and the requirement for the first moment's existence
gives the restriction $1<\alpha \leq 2$ \cite{L00a}. The author first solved
the fractional Schr\"{o}dinger equation for the infinite potential well, the
Bohr atom and introduced a fractional oscillator \cite{L00b}. Later he
derived the fractional Schr\"{o}dinger equation for three-dimensional motion
and solved it for the Bohr atom and a one-dimensional oscillator by means
of semiclassical approaches \cite{L02}.

Guo and Xu discussed the solutions of the fractional Schr\"{o}dinger equation
for the free particle, the infinite well, and a simple model for barrier
penetration, among other \textit{physical applications} \cite{GX06} and Dong
and Xu \cite{DX07} solved some other examples in the momentum representation.

Zoia et al \cite{ZRK07} addressed the problem of the boundary conditions in
fractional Laplacian equations and proposed a method for the accurate
calculation of eigenvalues and eigenfunctions that overcomes the difficulty
that had arisen in earlier approaches when $\alpha \rightarrow 2$. In
particular, they considered absorbing and free boundary conditions and took
advantage of the fact that the Laplacian equations exhibit exact solutions
when $\alpha $ is an even integer in order to test their method.

The purpose of this paper is to show that a simple collocation method is
suitable for the treatment of the fractional Schr\"{o}dinger equation. This
approach has already proved successful in standard quantum mechanics as well
as for several other physical problems \cite{ACF07,A07,A08,A09,AFSS09}. We
think that it is most convenient to have a method that applies to a wide
variety of problems and for this reason, in this paper, we propose the
ubiquitous collocation method based on little sinc functions (LSF) that
easily accommodate to a variety of boundary conditions \cite{ACF07}.

In Section \ref{sec:FQM} we outline the main features of fractional quantum
mechanics and discuss the fractional Laplacian operator in terms of well
known operator methods. In Section \ref{sec:method} we describe the collocation
method, develop the Fourier decomposition of the sampling functions to be used,
as well as calculate explicitly the effect of the fractional differentiation
operators on such sampling functions. In section \ref{sec:applications}, we apply
our collocation method to two fractional differentiation problems. Finally,
in section \ref{sec:conclusions} we present our conclusions.

\section{Fractional quantum mechanics}

\label{sec:FQM}

By means of the fractional path integral in which the L\'{e}vy motion
substitutes the Brownian one, Laskin \cite{L00a,L00b} derived the fractional
Schr\"{o}dinger equation
\begin{equation}
i\hbar \frac{\partial \psi }{\partial t}=\hat{H}_{\alpha }\psi ,\;\hat{H}%
_{\alpha }=-D_{\alpha }(\hbar \nabla )^{\alpha }+V(x) , \label{eq:TDSE_Laskin}
\end{equation}
where $D_{\alpha }$ is a generalized fractional diffusion coefficient. The
author proved that the fractional Schr\"{o}dinger operator $\hat{H}_{\alpha
} $ is Hermitian or self--adjoint\cite{L00a,L00b,L02}. The meaning of the
fractional derivative is clear from its effect upon a plane wave \cite
{L00a,L00b,L02}
\begin{equation}
\nabla ^{\alpha }e^{ikx}=|k|^{\alpha }e^{ikx} . \label{eq:frac_der_def_Laskin}
\end{equation}
Laskin \cite{L02} also considered a 3D generalization of the fractional
derivative:
\begin{equation}
-\nabla ^{\alpha }\rightarrow (-\Delta )^{\alpha /2},
\label{eq:frac_der_def_Laskin_3D}
\end{equation}
where $\Delta $ is the Laplacian operator.

On the other hand, Zoia et al \cite{L02} considered the alternative
definition
\begin{eqnarray}
\frac{\partial ^{\alpha }}{\partial |x|^{\alpha }}e^{iqx} &=&-|q|^{\alpha
}e^{iqx},  \nonumber \\
\frac{\partial ^{\alpha }}{\partial |x|^{\alpha }} &=&-(-\Delta )^{\alpha /2},
\label{eq:frac_der_def_Zoia}
\end{eqnarray}
that is consistent with the one above. Clearly, the effect of the
fractional derivative on an arbitrary function follows from the form of the
standard Fourier transform \cite{L00a,L00b,L00c,L02,GX06,DX07,ZRK07}.

In this paper we resort to the standard definition of the function of an
operator in order to define the fractional derivative. One advantage of this
definition, which is consistent with the one just outlined above, is that it
enables us to obtain several results without further proof. If $\hat{A}$ is
a Hermitian operator with a complete set of eigenvectors $\{|j>\}$,
\begin{equation}
\hat{A}|j>=a_{j}|j> , \label{eq:A|j>}
\end{equation}
then we define the function $f(\hat{A})$ by means of the spectral
decomposition \cite{RS80}
\begin{equation}
f(\hat{A})=\sum_{j}f(a_{j})|j><j| , \label{eq:f(A)}
\end{equation}
that makes sense if $f(x)$ is well defined for all $x=a_{j}$. The extension
to a continuous spectrum is straightforward. Notice that if $f(x)$ is real
then $f(\hat{A})$ is Hermitian. Besides, if $[\hat{A},\hat{B}]=0$ then,
obviously, $[f(\hat{A}),\hat{B}]=0$.

For example, on the whole line we have the standard definition in terms of
the\ Fourier transform \cite{L00a,L00b,L00c,L02,GX06,DX07,ZRK07}
\begin{equation}
(-\Delta )^{\alpha /2}\psi (x)=\frac{1}{2\pi }\int_{-\infty }^{\infty
}\int_{-\infty }^{\infty }|k|^{\alpha }e^{ik(x-y)}dk\,\psi (y)\,\,dy,
\label{eq:Fourier_int}
\end{equation}
and, in terms of the Fourier series for periodic boundary conditions,
\begin{equation}
(-\Delta )^{\alpha /2}\psi (x) = \frac{1}{2\pi }\int_{-\pi}^\pi
\sum_{n=-\infty }^{\infty } |n|^{\alpha }e^{in(x-y)}\psi (y)\,\,dy.
\label{eq:Fourier_series}
\end{equation}
In Section \ref{sec:method} we consider other functions and other boundary conditions.

In addition to being the basis for the numerical method proposed in this
paper, the well known general operator results outlined above enable us to
derive several conclusions without further proof. For example, if we choose
the domain of the operator $-\Delta $ so that it is Hermitian, then its
eigenvalues are real and positive. Its obvious consequence is that $(-\Delta
)^{\alpha /2}$ is also Hermitian \cite{L00a,L00b}.

It is well known that the Laplacian commutes with the operator that produces
the inversion transformation $\mathbf{r}\rightarrow -\mathbf{r}$; therefore,
$(-\Delta )^{\alpha /2}$ also commutes with the inversion operator \cite{L02}.

Every textbook on quantum mechanics shows that the nontrivial solutions to
the eigenvalue equation $(-\Delta \psi )(x)=-\psi ^{\prime \prime
}(x)=\lambda \psi (x)$ with boundary conditions $\psi (\pm a)=0$ are given
by
\begin{eqnarray}
\psi _{n}(x) &=&\frac{1}{\sqrt{a}}\sin \left[ \frac{n\pi (x+a)}{2a}\right] ,
\nonumber \\
\lambda &=&\lambda _{n}=\frac{n^{2}\pi ^{2}}{4a^{2}},\;n=1,2,\ldots
\label{eq:Part_box}
\end{eqnarray}
Therefore, it follows from the operator equations outlined above that
\begin{eqnarray}
\hat{T}_{\alpha }\psi _{n} &=&E_{n}^{\alpha }\psi _{n} , \nonumber \\
\hat{T}_{\alpha } &=&D_{\alpha }\left( -\hbar ^{2}\Delta \right)^{\alpha/2}
,\;E_{n}^{\alpha }=D_{\alpha }\left( \frac{\hbar n\pi }{2a}\right) ^{\alpha } .
\label{eq:Frac_part_box}
\end{eqnarray}
In other words, it is not necessary to solve the time--independent
fractional Schr\"{o}dinger equation explicitly \cite{L00b,GX06} because we
already know that the eigenfunctions are exactly those of the particle in an
infinite well.

As a final example, consider the time--dependent Schr\"{o}dinger equation
\begin{equation}
i\hbar \frac{\partial \psi (x,t)}{\partial t}=\hat{T}_{\alpha }\psi (x,t)
\label{eq:time_dep}
\end{equation}
with the boundary conditions $\psi (\pm a,t)=0$. Straightforward application
of the method of separation of variables leads to
\begin{equation}
\psi (x,t)=\sum_{n=1}^{\infty }e^{-itE_{n}^{\alpha }/\hbar }\psi
_{n}(x)\int_{-a}^{a}\psi _{n}(x)^{*}\psi (x,0)\,dx , \label{eq:psi(x,t)}
\end{equation}
that generalizes the result derived earlier for the same fractional model
\cite{L00b,GX06}.

Notice that present definition of the fractional derivative avoids the problem
of nonlocality of the Riesz derivative pointed out by Jeng et al \cite{JXHS08}.

\section{The method\label{sec:method}}

As stated above, our purpose is the application of a collocation method to
fractional differentiation problems on bounded intervals, with a variety of
boundary conditions \cite{AFSS09}.

This collocation method starts by sampling a function $f$ on $[-L,L]$ in a
finite uniform set of points $x_k$, and then interpolating by
\[
f(x) \approx \sum_{k=-N}^N f(x_k) s_k(N,L,x),
\]
where the sampling functions $s_k(N,L,x)$, called \emph{little sinc functions} (LSF).
They are defined as
\[
s_k(N,L,x) = \frac{\delta_N(x,x_k)}{\delta_N(x_k,x_k)},
\]
where $\delta_N(x,y) = \sum_{n=0}^N \phi_n(x)\phi_n(y)$ is
the partial sum kernel for a complete orthonormal set of functions
$\{\phi_n\}$ in $L^2([-L,L])$, suitable chosen according to given boundary
conditions \cite{AFSS09}.

Thus, for a given operator $T$, we define $Tf$ on $[-L,L]$ by
\[
Tf(x) = \sum_{k=-N}^N f(x_k) Ts_k(N,L,x),
\]
which then can be calculated by interpolating, in turn, the function
$Ts_k(N,L,x)$. As we are interested in the application of $T$ when given as
a function of differentiation, it is convenient to decompose each sampling
function $s_k(N,L,x)$ in exponential functions.

We therefore develop the Fourier decomposition for the sampling functions
$s_k(N,L,x)$, taken from the LSF sets in \cite{AFSS09}, in order
to apply fractional differentiation.

\subsection{Periodic boundary conditions}

When given periodic boundary conditions, $f(-L)=f(L)$,
we use as an orthonormal set the functions
\[
\psi_0(x) = \frac{1}{\sqrt{2L}}, \qquad
\psi_n(x) = \frac{1}{\sqrt L}\cos\Big(\frac{n\pi x}{L}\Big), \qquad
\phi_n(x) = \frac{1}{\sqrt L}\sin\Big(\frac{n\pi x}{L}\Big),
\]
with sampling points $x_k = \dfrac{2Lk}{2N+1}$, $|k|\le N$. We will
denote the corresponding set of sampling functions $s_k(N,L,x)$ as
$\LSF_1$.

In order to obtain the Fourier decomposition for $s_k(N,L,x)$, we start by
calculating
\[
\delta_N(x,y) = \sum_{n=0}^N \big( \psi_n(x)\psi_n(y) + \phi_n(x)\phi_n(y) \big)
= \frac{1}{2L} \sum_{n=-N}^N e^{i\frac{n\pi}{L}(x-y)}.
\]
One can verify that, for any $y\in[-L,L]$, $\delta_N(y,y) = \dfrac{2N+1}{2L}$
and hence
\[
\bar\delta_N(x,y) = \frac{1}{2N+1} \sum_{n=-N}^N e^{i\frac{n\pi}{L}(x-y)}.
\]
Therefore, since $x_k = \dfrac{2Lk}{2N+1}$, $|k|\le N$,
\[
s_k(N,L,x) = \bar\delta_N(x,x_k)
= \frac{1}{2N+1} \sum_{n=-N}^N e^{-i\frac{2nk\pi}{2N+1}} e^{i \frac{n\pi}{L} x}.
\]
Note that we can write
\[
\begin{split}
s_k(N,L,x) &= \frac{1}{2N+1} \sum_{n=-N}^N e^{-i\frac{2nk\pi}{2N+1}} e^{i \frac{2n\pi}{2L} x}
= \frac{1}{2N+1} \sum_{n=-2N}^{2N} \frac{1+(-1)^n}{2}
e^{-i\frac{nk\pi}{2N+1}} e^{i \frac{n\pi}{2L} x}\\
&= \frac{1}{4N+2} \sum_{n=-2N}^{2N} \big(1+(-1)^n\big)
e^{-i\frac{nk\pi}{2N+1}} e^{i \frac{n\pi}{2L} x}.
\end{split}
\]
We prefer the use of the last expression, as it resembles the formulas below.

\subsection{Dirichlet boundary conditions}

In the case of given Dirichlet boundary conditions, $f(-L)=f(L)=0$,
we use the complete orthonormal system
\[
\psi_n(x) = \frac{1}{\sqrt L} \cos\Big(\frac{(2n+1)\pi x}{2L}\Big),\qquad
\phi_n(x) = \frac{1}{\sqrt L} \sin\Big(\frac{(n+1)\pi x}{L}\Big),
\]
with sampling points $x_k = \dfrac{Lk}{N}$, $|k|\le N$. The corresponding
set of sampling functions will be denoted by $\LSF_2$.

Now, we have
\[
\begin{split}
\delta_N(x,x_k) &= \sum_{n=0}^{N-1} \big( \psi_n(x)\psi_n(x_k) + \phi_n(x)\phi_n(x_k) \big)
= \frac{1}{4L}
\sum_{n=-2N}^{2N} \big( e^{-i\frac{n\pi k}{2N}} - (-1)^n e^{i\frac{n\pi k}{2N}} \big)
e^{i\frac{n\pi}{2L}x},
\end{split}
\]
and note that $\delta_n(x_k,x_k) = \dfrac{N}{L}$. Thus
\[
s_k(N,L,x) = \frac{1}{4N}
\sum_{n=-2N}^{2N} \big( e^{-i\frac{n\pi k}{2N}} - (-1)^n e^{i\frac{n\pi k}{2N}} \big)
e^{i\frac{n\pi}{2L}x}.
\]

We observe that
\[
e^{-i\frac{n\pi k}{2N}} - (-1)^n e^{i\frac{n\pi k}{2N}} =
e^{i\frac{n\pi}{2}} \big( e^{-i\frac{n\pi}{2}}e^{-i\frac{n\pi k}{2N}} -
e^{i\frac{n\pi}{2}} e^{i\frac{n\pi k}{2N}} \big) =
2 i^{n-1} \sin\Big[\Big( \frac{1}{2} + \frac{k}{2N} \Big)\pi n\big],
\]
so we obtain
\[
s_k(N,L,x) = \frac{1}{2N}
\sum_{n=-2N}^{2N} i^{n-1} \sin\Big[\Big( \frac{1}{2} + \frac{k}{2N} \Big)\pi n\big]
e^{i\frac{n\pi}{2L}x}.
\]

\subsection{Antiperiodic boundary conditions}

When given antiperiodic boundary conditions, $f(-L) = -f(L)$,
we take as a complete orthonormal system
\[
\psi_n(x) = \frac{1}{\sqrt L} \cos\Big( \frac{2n+1}{2L} \pi x \Big), \qquad
\phi_n(x) = \frac{1}{\sqrt L} \sin\Big( \frac{2n+1}{2L} \pi x \Big),
\]
with sampling points $x_k=\dfrac{Lk}{N}$, $|k|\le N$. We will denote the
corresponding set of sampling functions by $\LSF_3$.

This time we have
\[
\begin{split}
\delta_N(x,y) &= \sum_{n=0}^{N-1} \big( \psi_n(x)\psi_n(y) + \phi_n(x)\phi_n(y) \big)
= \frac{1}{2L} \sum_{n=-N}^{N-1} e^{i\frac{2n+1}{2L}\pi (x-y)}\\
&=
\frac{1}{4L} \sum_{n=-2N}^{2N} \big( 1 - (-1)^n \big) e^{i\frac{n\pi}{2L} (x-y)},
\end{split}
\]
since we are adding only on odd numbers. As $\delta_N(y,y) = \dfrac{N}{L}$ for every
$y\in[-L,L]$, we have
\[
s_k(N,L,x) = \frac{1}{4N}
\sum_{n=-2N}^{2N} \big( 1 - (-1)^n \big) e^{-i\frac{nk\pi}{2N}} e^{i\frac{n\pi}{2L} x}
\]

\subsection{Neumann boundary conditions}

In the case of given Neumann boundary conditions, $f'(-L)=f'(L)=0$,
we consider the complete orthonormal system
\[
\psi_0(x) = \frac{1}{\sqrt{2L}}, \qquad
\psi_n(x) = \frac{1}{\sqrt L} \cos\Big( \frac{n}{L} \pi x \Big), \qquad n=1,2,\ldots,
\]
\[
\phi_n(x) = \frac{1}{\sqrt L} \cos\Big( \frac{2n+1}{2L} \pi x \Big),
\qquad n=0,1,2,\ldots,
\]
with sampling points $x_k=\dfrac{2Lk}{2N+1}$, $|k|\le N$. We denote the corresponding
sampling function set by $\LSF_4$.

This time we have
\[
\begin{split}
\delta_N(x,x_k) &= \sum_{n=0}^N \big( \psi_n(x)\psi_n(x_k) + \phi_n(x)\phi_n(x_k) \big)\\
&= \frac{1}{4L} \sum_{n=-2N}^{2N}
\big( e^{-i\frac{n\pi k}{2N+1}} + (-1)^n e^{i\frac{n\pi k}{2N+1}} \big)
e^{i\frac{n\pi}{2L}x},
\end{split}
\]
and $\delta_N(x_k,x_k) = \dfrac{2N+1}{2L}$. Thus, we obtain
\[
s_k(N,L,x) = \frac{1}{2(2N+1)}
\sum_{n=-2N}^{2N} \big( e^{-i\frac{n\pi k}{2N+1}} +
(-1)^n e^{i\frac{n\pi k}{2N+1}} \big) e^{i\frac{n\pi}{2L}x}.
\]
Using the identity
\[
\begin{split}
e^{-i\frac{n\pi k}{2N+1}} + (-1)^n e^{i\frac{n\pi k}{2N+1}} &=
e^{i\frac{n\pi}{2}} ( e^{-i\frac{n\pi}{2}} e^{-i\frac{n\pi k}{2N+1}} +
e^{i\frac{n\pi}{2}} e^{i\frac{n\pi k}{2N+1}} ) =
2i^n \cos \Big[\Big(\frac{1}{2} + \frac{k}{2N+1} \Big) n\pi\Big],
\end{split}
\]
we have
\[
s_k(N,L,x) = \frac{1}{2N+1}
\sum_{n=-2N}^{2N} i^n \cos \Big[\Big(\frac{1}{2} + \frac{k}{2N+1} \Big) n\pi\Big]
e^{i\frac{n\pi}{2L}x}.
\]

As a summary, we note that we have written all the sampling functions in the form
\[
s_k(N,L,x) = \sum_{n=-2N}^{2N} C_n(k,N) e^{i\frac{n\pi}{2L} x},
\]
where the coefficients $C_n(k,N)$ are given by
\[
C_n(k,N) =
\begin{cases}
\dfrac{1+(-1)^n}{2(2N+1)} e^{-i\frac{nk\pi}{2N+1}} & \LSF_1\\
\dfrac{i^{n-1}\sin(\frac{1}{2}+\frac{k}{2N})n\pi}{2N} & \LSF_2\\
\dfrac{1-(-1)^n}{4N} e^{-i\frac{nk\pi}{2N}} & \LSF_3\\
\dfrac{i^n\cos(\frac{1}{2}+\frac{k}{2N+1})n\pi}{2N+1} & \LSF_4.
\end{cases}
\]
Note that these coefficients do not depend on the length of the interval $[-L,L]$.

\subsection{Differential operators}

We now note that, after taking the derivative of each $s_k$, we obtain
\[
\frac{d}{dx}s_k(N,L,x) = \sum_{n=-2N}^{2N} C_n(k,N) \Big(i\frac{n\pi}{2L}\Big)
e^{i\frac{n\pi}{2L} x},
\]
so, if $\hat p$ is the momentum operator $\hat p=-i\dfrac{d}{dx}$, we obtain
\[
\hat p s_k(N,L,x) = \sum_{n=-2N}^{2N} C_n(k,N) \Big(\frac{n\pi}{2L}\Big)
e^{i\frac{n\pi}{2L} x}.
\]
We then define, for a given function $m$, the operator $m(\hat p)$ on the functions $s_k$ by
\[
m(\hat p) s_k(N,L,x) = \sum_{n=-2N}^{2N} C_n(k,N) m\Big(\frac{n\pi}{2L}\Big)
e^{i\frac{n\pi}{2L} x},
\]
i.e., we define $m(\hat p)$ through the spectrum of $\hat p$.

We are particularly interested in the case of the fractional operator $(-\Delta)^{\alpha/2}$, where
$\Delta$ is the Laplacian $\Delta = \dfrac{d^2}{dx^2} = -\hat p^2$. Thus, we have
$(-\Delta)^{\alpha/2} = |\hat p|^\alpha$, and
\[
(-\Delta)^{\alpha/2} s_k(N,L,x) =
\sum_{n=-2N}^{2N} C_n(k,N) \Big(\frac{|n|\pi}{2L}\Big)^\alpha e^{i\frac{n\pi}{2L} x}.
\]

In order to interpolate the resulting functions $(-\Delta)^{\alpha/2}s_k$,
we calculate its value in the sampling points $x_j$, i.e.
\[
(-\Delta)^{\alpha/2} s_k(N,L,x_j) =
\sum_{n=-2N}^{2N} C_n(k,N) \Big(\frac{|n|\pi}{2L}\Big)^\alpha e^{i\frac{n\pi}{2L} x_j}.
\]
We do this explicitly for each of the sampling function sets from above.

\begin{enumerate}
\item[($\LSF_1$)] We first calculate it for the set $\LSF_1$. In this case
$x_j = \dfrac{2Lj}{2N+1}$ and
\[
C_n(k,N) = \frac{1+(-1)^n}{2(2N+1)} e^{-i\frac{nk\pi}{2N+1}},
\]
and thus
\[
C_n(k,N) \Big(\frac{|n|\pi}{2L}\Big)^\alpha e^{i\frac{n\pi}{2L} x_j} =
\Big(\frac{|n|\pi}{2L}\Big)^\alpha \frac{1+(-1)^n}{2(2N+1)}
e^{-\frac{in(k-j)\pi}{2N+1}},
\]
so
\[
\begin{split}
(-\Delta)^{\alpha/2} s_k(N,L,x_j) &= \sum_{n=-2N}^{2N}
\Big(\frac{|n|\pi}{2L}\Big)^\alpha \frac{1+(-1)^n}{2(2N+1)}
e^{-\frac{in(k-j)\pi}{2N+1}} \\ &=
\frac{2}{2N+1} \sum_{n=1}^N \Big(\frac{n\pi}{L}\Big)^\alpha
\cos \frac{2n(k-j)\pi}{2N+1}.
\end{split}
\]
\item[($\LSF_2$)] For this set, $x_j = \dfrac{Lj}{N}$ and
\[
C_n(k,N) = \frac{i^{n-1}}{2N}\sin\Big(\frac{1}{2}+\frac{k}{2N}\Big)n\pi,
\]
so
\[
C_n(k,N) \Big(\frac{|n|\pi}{2L}\Big)^\alpha e^{i\frac{n\pi}{2L} x_j} =
\Big(\frac{|n|\pi}{2L}\Big)^\alpha \frac{i^n}{2Ni}
\sin\Big[\Big(\frac{1}{2}+\frac{k}{2N}\Big)n\pi\Big] e^{i\frac{nj\pi}{2N}}.
\]
Thus we obtain
\[
\begin{split}
(-\Delta)^{\alpha/2} s_k(N,L,x_j) &= \sum_{n=-2N}^{2N}
\Big(\frac{|n|\pi}{2L}\Big)^\alpha \frac{i^n}{2Ni}
\sin\Big[\Big(\frac{1}{2}+\frac{k}{2N}\Big)n\pi\Big] e^{i\frac{nj\pi}{2N}}\\
&=
\frac{1}{2N} \sum_{n=1}^{2N} \Big(\frac{n\pi}{2L}\Big)^\alpha \Big(
\cos\frac{n(k-j)\pi}{2N} - (-1)^n \cos\frac{n(k+j)\pi}{2N} \Big).
\end{split}
\]
\item[($\LSF_3$)] For this set, $x_j = \dfrac{Lj}{N}$ and
\[
C_n(k,N) = \frac{1-(-1)^n}{4N} e^{-i\frac{nk\pi}{2N}},
\]
so
\[
C_n(k,N) \Big(\frac{|n|\pi}{2L}\Big)^\alpha e^{i\frac{n\pi}{2L} x_j} =
\Big( \frac{|n|\pi}{2L} \Big)^\alpha \frac{1-(-1)^n}{4N}e^{-\frac{in(k-j)\pi}{2N}},
\]
and therefore
\[
\begin{split}
(-\Delta)^{\alpha/2} s_k(N,L,x_j) &= \sum_{n=-2N}^{2N}
\Big( \frac{|n|\pi}{2L} \Big)^\alpha \frac{1-(-1)^n}{4N}e^{-\frac{in(k-j)\pi}{2N}} \\
&=
\frac{1}{N} \sum_{n=1}^N \Big( \frac{(2n-1)\pi}{2L}\Big)^\alpha
\cos \frac{(2n-1)(k-j)\pi}{2N}.
\end{split}
\]
\item[($\LSF_4$)] Finally, for the set $\LSF_4$, the sampling points are given
by $x_j = \dfrac{2Lj}{2N+1}$ and
\[
C_n(k,N) = \frac{i^n}{2N+1}\cos\Big(\frac{1}{2}+\frac{k}{2N+1}\Big)n\pi,
\]
so we have
\[
C_n(k,N) \Big(\frac{|n|\pi}{2L}\Big)^\alpha e^{i\frac{n\pi}{2L} x_j} =
\Big(\frac{|n|\pi}{2L}\Big)^\alpha
\frac{i^n}{2N+1}\cos\Big[\Big(\frac{1}{2}+\frac{k}{2N+1}\Big)n\pi\Big]
e^{\frac{inj\pi}{2N+1}}.
\]
Thus
\[
\begin{split}
(-\Delta)^{\alpha/2} s_k(N,L,x_j) &= \sum_{n=-2N}^{2N}
\Big(\frac{|n|\pi}{2L}\Big)^\alpha
\frac{i^n}{2N+1}\cos\Big[\Big(\frac{1}{2}+\frac{k}{2N+1}\Big)n\pi\Big]
e^{\frac{inj\pi}{2N+1}}\\
&= \frac{1}{2N+1} \sum_{n=1}^{2N} \Big(\frac{n\pi}{2L} \Big)^\alpha \Big(
\cos\frac{n(k-j)\pi}{2N+1} + (-1)^n \cos\frac{n(k+j)\pi}{2N+1} \Big).
\end{split}
\]
\end{enumerate}

\subsection{Collocation}

As described above, we approximate a function $f$ on $[-L,L]$ through
interpolation from the sampling points $x_k$ by means of
\[
f(x) \approx \sum_{k=-N}^N f(x_k) s_k(N,L,x).
\]
Thus the action of $m(\hat p)$ is defined by
\[
\begin{split}
m(\hat p)f(x) &= \sum_{k=-N}^N f(x_k) m(\hat p)s_k(N,L,x) \\
&= \sum_{k=-N}^N f(x_k) \sum_{j=-N}^N m(\hat p)s_k(N,L,x_j) s_j(N,L,x)\\
&= \sum_{j=-N}^N \bigg( \sum_{k=-N}^N f(x_k) \sum_{n=-2N}^{2N} C_n(k,N) m\Big(\frac{n\pi}{2L}\Big)
e^{i\frac{n\pi}{2L} x_j} \bigg) s_j(N,L,x).
\end{split}
\]

Note that we can view this as the action of the matrix
\[
M^N\hat p = [M^N\hat p]_{kj}, \qquad -N\le k,j\le N,
\]
on the vectors $f(x_k)$, $-N\le k\le N$, where
the matrix entries are given by
\[
[M^N\hat p]_{kj} = \sum_{n=-2N}^{2N} C_n(k,N) m\Big(\frac{n\pi}{2L}\Big)
e^{i\frac{n\pi}{2L} x_j}.
\]
Note that we have explicitly calculated these coefficients for
$(-\Delta)^{\alpha/2} = |\hat p|^\alpha$ in the previous section.

\section{Applications}\label{sec:applications}

In this section we apply the present collocation method to the fractional versions
of the anharmonic oscillator and the Mathieu equation.

\subsection{Fractional oscillators}

Our first example is the fractional oscillator
\beq
\hat{H} = D_\alpha (-\hbar^2 \Delta)^{\alpha/2} + q^2 |\vec{r}|^{\beta} ,
\label{fho}
\eeq
studied by Laskin \cite{L00a} by means of the semiclassical WKB approach.
In the one-dimensional case Laskin obtained the following approximate analytical
expression for the energies:
\beq
E_n = \Big( \frac{\pi \hbar \beta D_\alpha^{1/\alpha} q^{2/\beta}}{2 B(1/\beta, 1/\alpha+1)}
\Big)^{\frac{\alpha \beta}{\alpha+\beta}} \
\Big( n + \frac{1}{2}\Big)^\frac{\alpha \beta}{\alpha+\beta} ,
\label{wkb}
\eeq
where $B(1/\beta, 1/\alpha+1)$ is the beta function.

This problem is suitable for illustrating the application of the collocation method
described above. For concreteness we choose a dimensionless model with
$D_\alpha = q = \hbar = 1$ and resort to the set $LSF_2$ because the Dirichlet
boundary conditions are suitable for this problem. We obtain reasonably accurate
results with $N=50$, that corresponds to  $99$ sampling points.

The LSF are defined on an interval
$|x| \leq L$. The unphysical parameter $L$ may lead to inaccurate results if it is
not chosen properly. If it is too small the wave function will decay too rapidly. If,
on the other hand, $L$ is too large then we would need an unnecessarily large number
$N$ of sampling points in order to have sufficiently accurate results. In order to
get a reasonable balance between those parameters of the LSF method we resort to the
strategy followed in earlier applications of the  collocation approach
based on sinc functions \cite{A06}  and then extended to LSF \cite{ACF07}.
Since the trace of the Hamiltonian matrix is invariant under unitary transformations,
and the actual eigenvalues are independent of $L$, then it is reasonable to choose the
value of $L$ close to a stationary point. This principle of minimal sensitivity
(PMS)~\cite{Ste81} gives the optimal value of $L$ for a given value of $N$. In
the present case the stationary point is a minimum that we will call $L_{PMS}$
from now on.

In Table \ref{table-1} we report the three lowest eigenvalues of the fractional
harmonic oscillator ($\beta=2$) with $\alpha = 3/2$ for grids of
varying size. We appreciate that the value of $L_{PMS}$ shown in the second
column grows with $N$ in agreement with the argument above. The last row shows
the energies obtained by application of exactly the same approach to the
Schr\"odinger equation in the momentum representation, where no fractional derivatives
are present, for a much finer grid. Notice that the rate of convergence of the present
collocation method depends on $\alpha$. In order to appreciate this point more clearly
we may compare present results for $N=100$, which exhibit only $4$ exact digits,
with identical calculation for the standard harmonic oscillator ($\alpha = 2$),
where just $N=10$ enables us to obtain $E_0^{PMS} \approx 0.9999999999991$ that is
about $10^{-13}$ off the exact value $E_0^\text{exact}=1$.

\begin{table}[!htb]
\caption{Three lowest energy eigenvalues of the fractional harmonic oscillator
($\beta=2$) with $\alpha = 3/2$. The last row shows the results for the Schr\"odinger
equation in momentum space.}
\label{table-1}
\begin{tabular}{|c||c||c|c|c|}
    \hline
$N$ &  $L_{PMS}$ & $E_0$    &   $E_{1}$  & $E_{2}$   \\
    \hline \hline
$10$   & 4.366 & 1.010039766 & 2.710385528 & 4.18329885 \\
$20$   & 5.797 & 1.005291363 & 2.708645561 & 4.17935372 \\
$30$   & 6.866 & 1.003815977 & 2.708337656 & 4.17844614 \\
$40$   & 7.751 & 1.003106441 & 2.708230888 & 4.17805574 \\
$50$   & 8.518 & 1.002691899 & 2.708181518 & 4.17784097 \\
$60$   & 9.202 & 1.002421030 & 2.708154647 & 4.17770589 \\
$70$   & 9.825 & 1.002230636 & 2.708138397 & 4.17761342 \\
$80$   & 10.40 & 1.002089737 & 2.708127815 & 4.17754632 \\
$90$   & 10.93 & 1.001981392 & 2.708120532 & 4.17749550 \\
$100$  & 11.43 & 1.001895574 & 2.708115301 & 4.17745573 \\
\hline \hline
$500$  & 69.11 & 1.000989809 & 2.708093424 & 4.17706229 \\
\hline
\end{tabular}
\end{table}

\begin{figure}[ht]
\bigskip\bigskip
\begin{center}
\includegraphics[width=9cm]{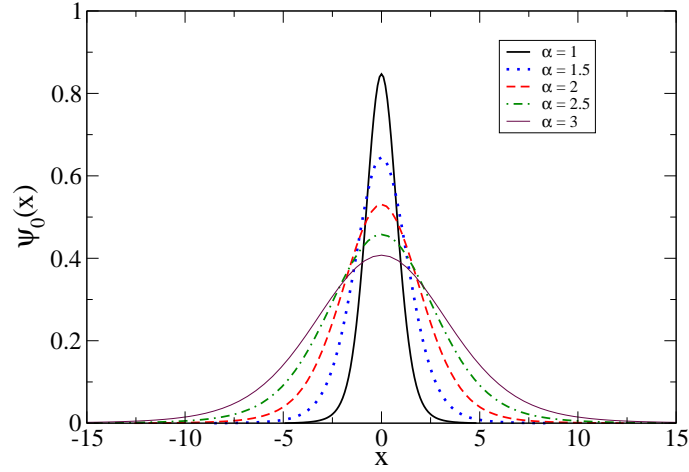}
\end{center}
\par
\bigskip
\caption{Ground state of the fractional harmonic oscillator
for different values of $\alpha$, obtained with $N=50$.}
\label{fig:1}
\end{figure}

In Figure \ref{fig:1} we have plotted the wave function of the ground state of
the fractional harmonic oscillator for different values of $\alpha$ and using a
grid with $N=50$. The case $\alpha = 2$ is the exact Gaussian
wave function of the standard harmonic oscillator.

\begin{figure}[ht]
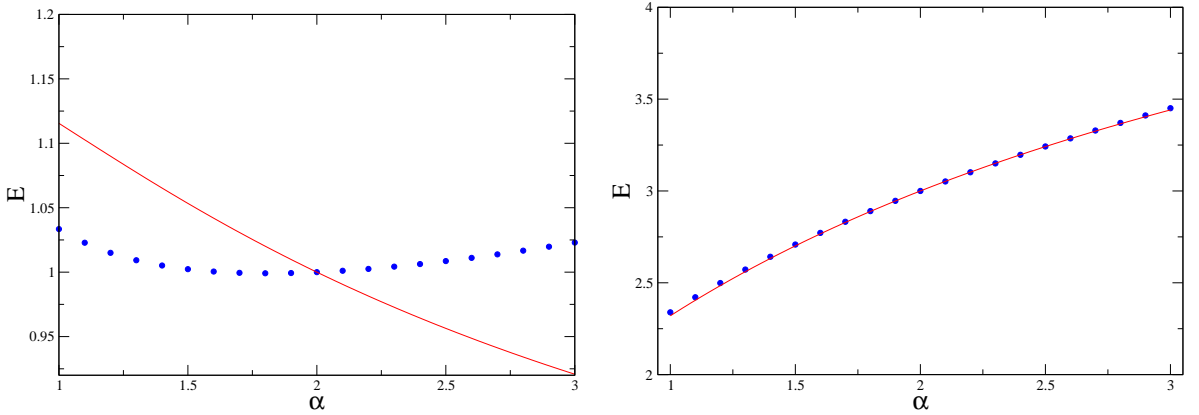

\bigskip\bigskip
\begin{center}
\includegraphics[width=3.in]{fig-2A.eps} \ \ \
\includegraphics[width=3.in]{fig-2B.eps}
\end{center}
\par
\bigskip
\caption{Energies of the first two states of the fractional harmonic oscillator
as functions of $\alpha$. Dots and lines mark present results for $N=60$ and the
WKB ones, respectively.}
\label{fig:1b}
\end{figure}

In Figure \ref{fig:1b} we have plotted the energies of
the first two states of the fractional harmonic oscillator for different values
of $\alpha$, using a grid with $N=60$. It also shows the results given by
the WKB formula derived by Laskin \cite{L02}. The agreement for the first excited state
is remarkable, and the large deviation for the ground state is not surprising because
the WKB method is expected to be valid for sufficiently large quantum numbers.

\begin{figure}[ht]
\bigskip\bigskip
\begin{center}
\includegraphics[width=9cm]{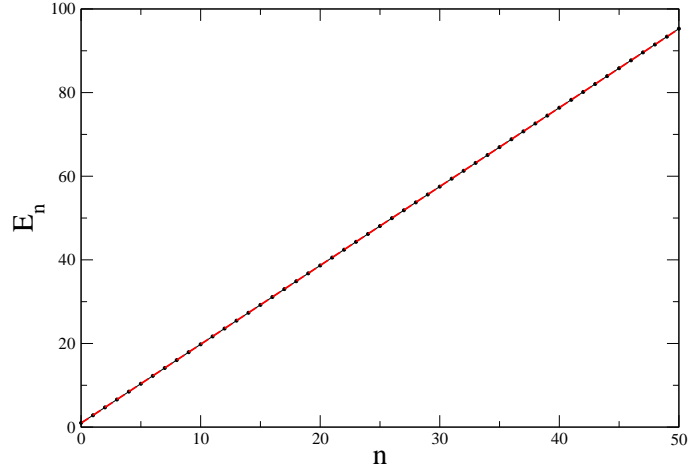}
\end{center}
\par
\bigskip
\caption{Energies of the fractional quartic anharmonic oscillator with $\alpha=4/3$,
calculated using $LSF_2$ with $N=50$ (points). The solid line is the least--squares
fit $E_n = 0.941 + 1.886\,n$.}
\label{fig:2}
\end{figure}

\begin{figure}[ht]
\bigskip\bigskip
\begin{center}
\includegraphics[width=9cm]{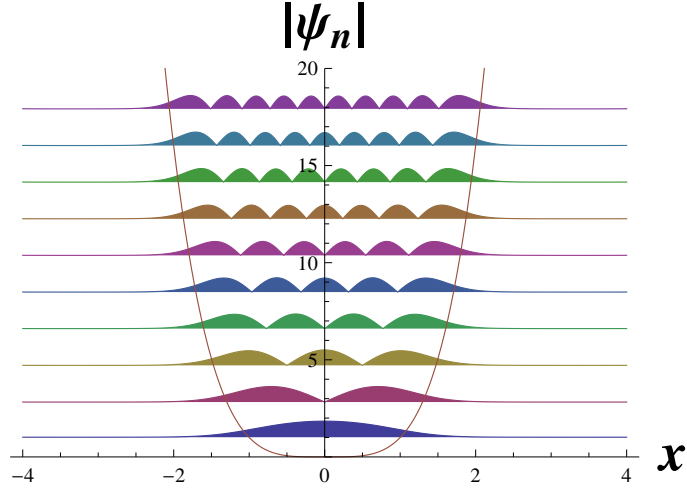}
\end{center}
\par
\bigskip
\caption{Absolute values of the wave functions (absolute value) of the fractional
quartic anharmonic oscillator with $\alpha=4/3$ and $\beta=4$
calculated using $LSF_2$ with $N=50$.}
\label{fig:3}
\end{figure}

Finally, we show results for the anharmonic oscillator
$V(x) = x^4$ with a fractional Laplacian corresponding to $\alpha = 4/3$.
We choose this particular example because the WKB formula predicts its energy spectrum
to be evenly spaced, like the standard harmonic oscillator. Figures \ref{fig:2}
and \ref{fig:3} display the energies of this fractional anharmonic oscillator and
the absolute value of the first ten wave functions. The spectrum follows the straight
line  $E_n = 0.941 + 1.886\,n$ in good agreement with the WKB estimate
\[
E_n = \frac{2\pi}{B(1/4, 3/4 + 1)}\Big( n + \frac{1}{2}\Big)
= \frac{2\pi}{\Gamma(1/4)\Gamma(7/4)}\Big( n + \frac{1}{2}\Big)
\approx 1.88562 \times \Big( n + \frac{1}{2}\Big).
\]

\subsection{Fractional Mathieu equation}

In order to illustrate the application of the present collocation method to a
problem with other boundary conditions than those discussed earlier, we
consider the fractional extension to the well known Mathieu equation \cite{AbSteg}
\begin{equation}\label{Math-eq}
\frac{d^{2}y}{dz^{2}}+(a-2q\cos 2z)y=0 ,
\end{equation}

given by
\begin{equation}\label{frac-Math}
(-\Delta )^{\alpha /2}y - (a-2q\cos 2z)y=0 ,
\end{equation}
which reduces to \eqref{Math-eq} when $\alpha =2$. We
resort to the set of periodic boundary conditions that apply to the
little sinc functions $\LSF_1$. In this case $L$ is fixed to $L=\pi$.

We follow the standard notation
and denote by $a_i$, $i=0,1,2,\ldots$  \cite{AbSteg},
the eigenvalues with even periodic functions, and by $b_i$, $i=1,2,\ldots$, the
eigenvalues corresponding to odd periodic functions. Note that, in the case
of $q=0$, $a_0=0$ and the eigenfunction is constant.

Table \ref{table-2} shows the convergence rate of our method for the lowest eigenvalue
$a_0$ of the fractional Mathieu equation with $\alpha =1$, $\alpha =3/2$,
and $q=1$. The
rate of convergence for the Mathieu equation is considerably larger than the
one discussed above for the fractional harmonic oscillator. In both cases it
increases as $\alpha$ approaches the ordinary value $\alpha =2$.

\begin{table}[ht]
\caption{Rate of convergence for the eigenvalue ($a_0$) of the fractional
Mathieu equation.}
\label{table-2}
{\footnotesize \begin{tabular}{|c||c|c|}
\hline
$N$ & $\alpha=1$ & $\alpha=3/2$ \\ \hline\hline
$10$ & -0.78002010749909950036806303597771247593586508305415 &
-0.60337681905510495225302969295010866313871477789813 \\
$20$ & -0.78002010679715466707531498072556654488485913105173 &
-0.60337681905490085108768066745917280899338056699081 \\
$30$ & -0.78002010679715466707518738897487326807118147127619 &
-0.60337681905490085108768066745913456676371052157098 \\
$40$ & -0.78002010679715466707518738897487326774021967916894 &
-0.60337681905490085108768066745913456676371052157098 \\ \hline\hline
exact & -0.78002010679715466707518738897487326774021967916894 &
-0.60337681905490085108768066745913456676371052157098 \\ \hline
\end{tabular}}
\end{table}

Table \ref{table-3} shows the four lowest eigenvalues of the fractional Mathieu
equation for different values of $\alpha$, again with $q=1$, as well as the symmetry
and periodicity of each wavefunction. They agree with the known results when
$\alpha=2$ \cite{AbSteg}.

\begin{table}[th]
\caption{Eigenvalues $a_0, b_1, a_1, b_2$ of the fractional
Mathieu equation with $q=1$ and different values of $\alpha$ ($N=50$).}
\label{table-3}
\begin{tabular}{|c||c|c|c|c|}
\hline
$\alpha$ & $a_0$ & $b_1$ & $a_1$ & $b_2$ \\ \hline\hline
$1$ & -0.78002010679715466708 & -0.31981501215423234713 &
1.2959422293970261239 & 1.5491290256879243036 \\
$\frac{3}{2}$ & -0.60337681905490085109 & -0.18880108186701679596 &
1.7046089276653617549 & 2.6389530962188063857 \\
$2$ & -0.45513860410741354823 & -0.11024881699209516991 &
1.8591080725143634723 & 3.9170247729984711867 \\
$\frac{5}{2}$ & -0.33549116582363455500 & -0.06396091681659914089 &
1.9267035413113906794 & 5.6189308675791269007 \\
$3$ & -0.24308662756250760871 & -0.03699729990815279808 &
1.9600508496994480694 & 7.9821470161415594702 \\ \hline
Symmetry & even & odd & even & odd \\ \hline
Period & 2 $\pi$ & $\pi$ & 2 $\pi$ & $\pi$ \\ \hline
\end{tabular}
\end{table}

\begin{figure}[th]
\begin{center}

\includegraphics[width=2in]{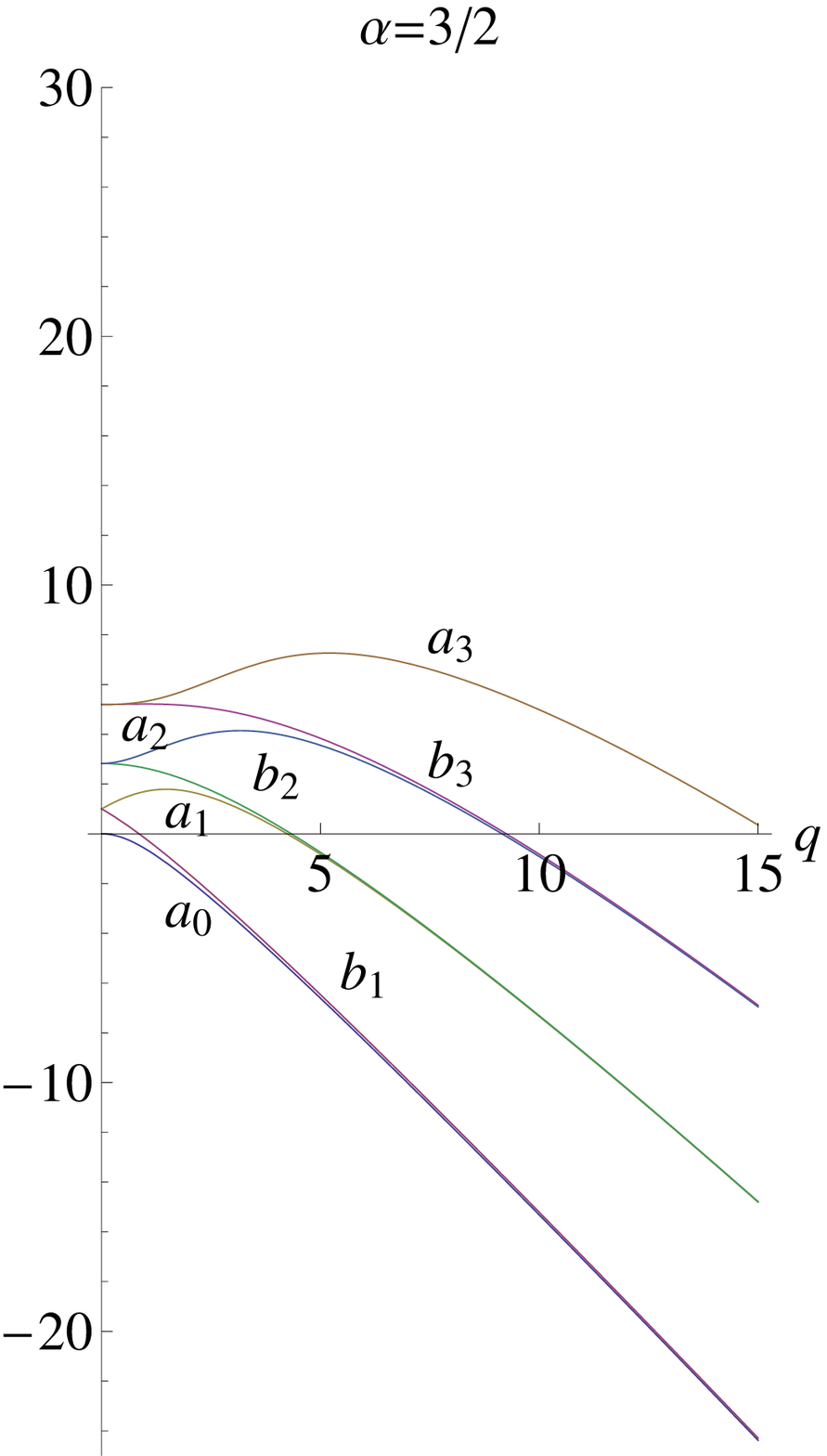}
\includegraphics[width=2in]{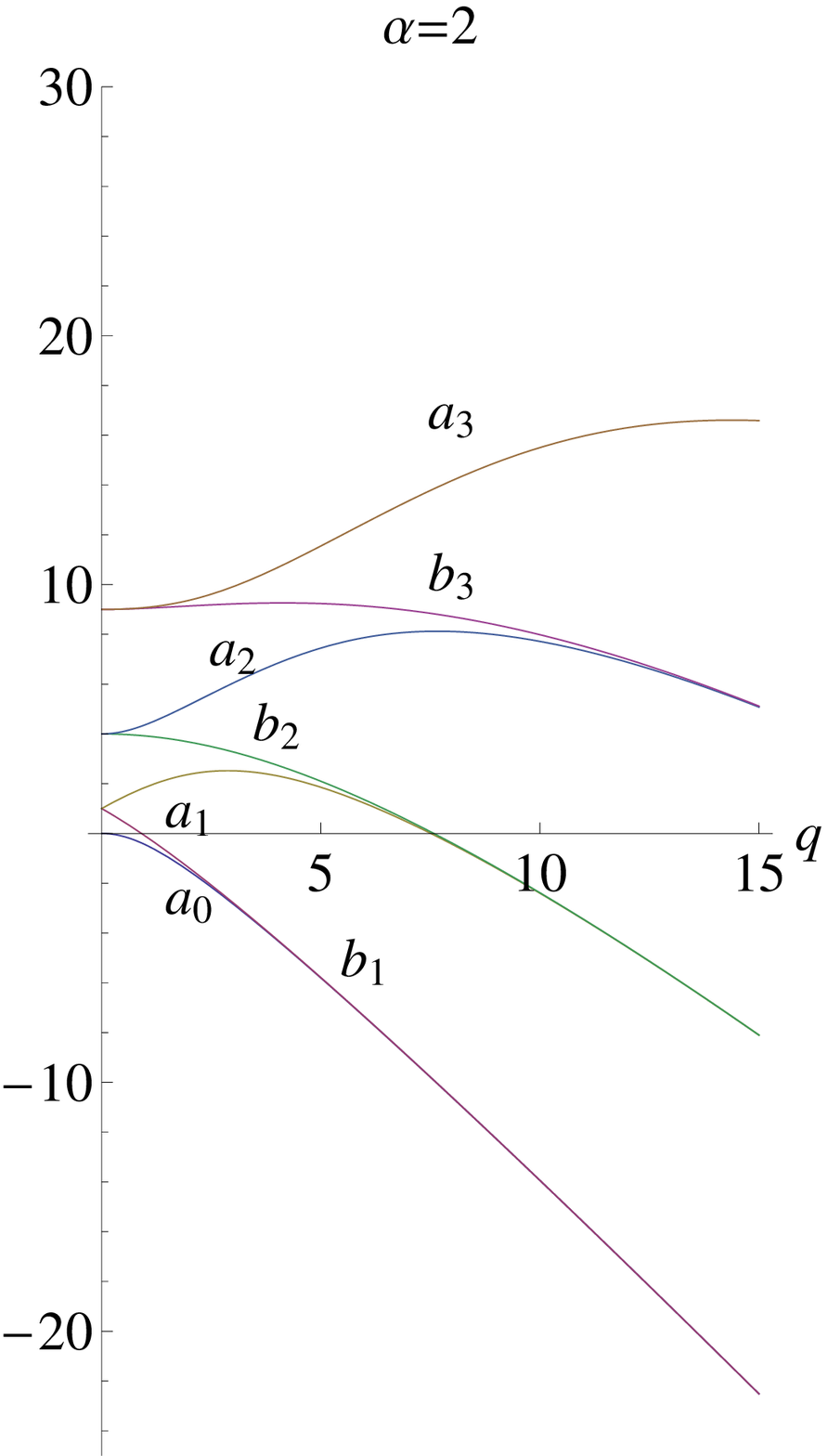}
\includegraphics[width=2in]{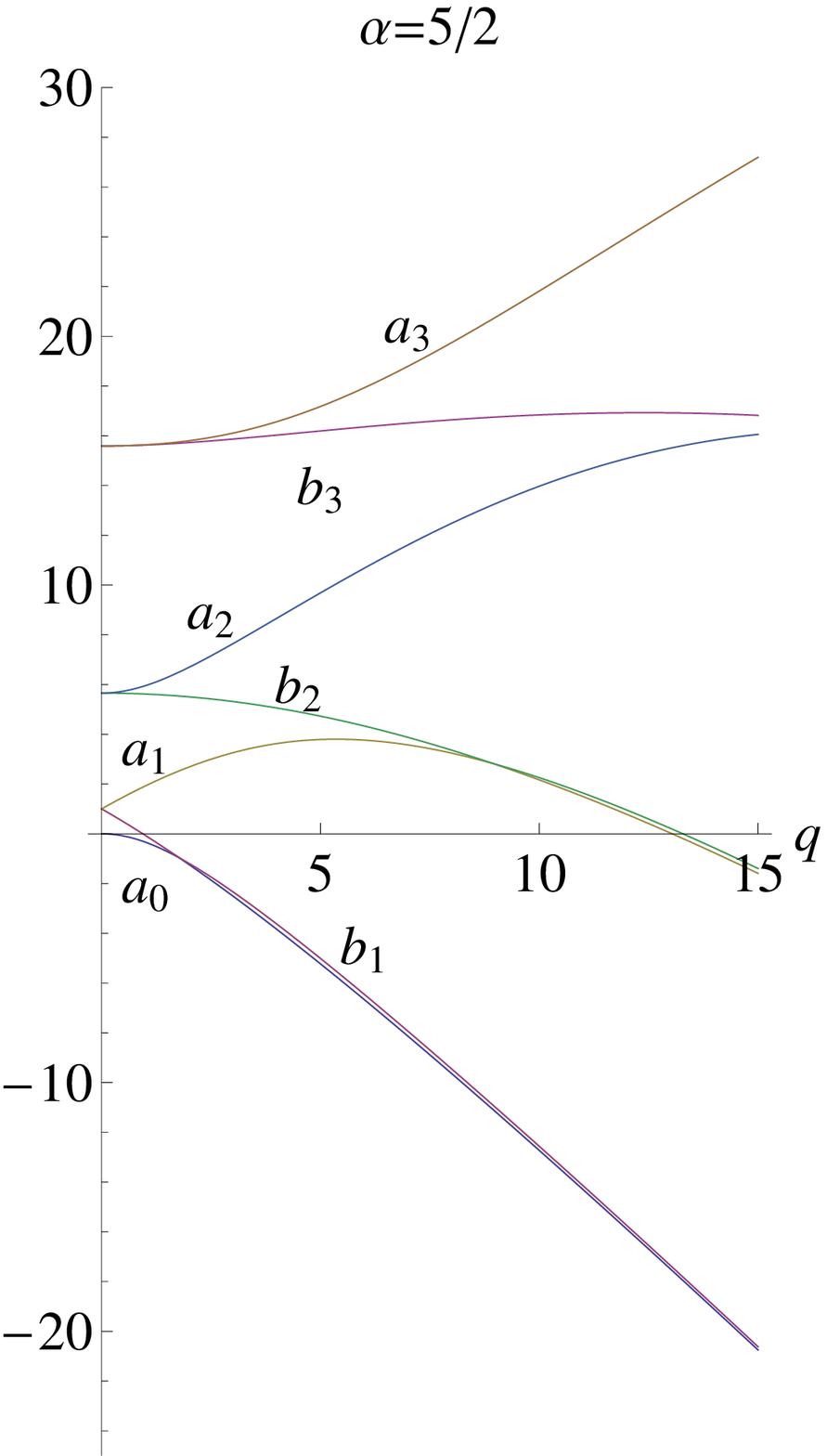}
\end{center}
\caption{Eigenvalues of the fractional Mathieu equation
for three values of $\alpha$.}
\label{fig:4}
\end{figure}

Figure \ref{fig:4} shows the eigenvalues
$a_i$, $i=0,1,2,3$, $b_i$, $i=1,2,3$, for the fractional Mathieu equation
with $\alpha =2, 3/2, 5/2$ obtained with the periodic set $\LSF_{1}$ and $N=30$
for $q\in[0,15]$ (cf. \cite[Figure 20.1]{AbSteg}).
Observe that the overall pattern of the behavior of these
eigenvalues with $q$ is similar for all values of $\alpha$. This fact is hardly
surprising because the degeneracy at $q=0$ and $q\rightarrow \infty$ is a consequence of
the form of the periodic potential.

In this extremely simple case it may be more practical to resort to the
standard method \cite{AbSteg} because the calculation of the matrix elements
of $\cos 2z$ does not offer any difficulty. We have chosen this example
simply as a test; however, one will notice the advantage of the collocation
method in the case of an arbitrary periodic potential $V(z)$ where the
calculation of the matrix elements may not be so simple.

\section{Conclusions\label{sec:conclusions}}

We have devised a numerical scheme based on collocation which allows one to solve
the fractional Schr\"odinger equation on a
uniform grid. We have applied this method to obtain accurate energies
and wave functions of a fractional
harmonic oscillator and compared the former with the WKB ones~\cite{L02}.
Our results confirm that the WKB approach yields reasonable results for excited states.
In the case of the fractional harmonic oscillator the WKB formula does not give the
exact result for all quantum numbers as in the ordinary case $\alpha=2$.

We have also studied a fractional anharmonic oscillator  for which
the WKB formula predicts equally spaced levels like the standard harmonic oscillator.
Our accurate results confirm this prediction beyond any doubt.
Finally we have solved a fractional Mathieu equation, with periodic boundary conditions,
and obtained the eigenvalues for different values of $\alpha$ and potential strength.

One of the main advantages of the collocation methods in general is that they bypass
the problem of calculating the matrix elements of the potential. This feature is most
welcome when the potential--energy function is rather complicated. This is not the case
of the models chosen here because we have been mainly interested in the discussion
of the fractional kinetic energy.

\begin{acknowledgments}
F. M. Fern\'andez acknowledges support of the Universidad de Colima through the
PIFI program.
\end{acknowledgments}

\end{document}